%% file: ms_astroph.tex
\documentclass[numberedappendix]{emulateapj}
\usepackage{apjfonts}
\usepackage{amsmath}
\usepackage{amsbsy}
\usepackage{amssymb}
\usepackage{epsfig}
\usepackage{natbib}

\newcommand{\kms}{\,{\rm km\,s^{-1}}}
\newcommand{\au}{\,{\rm AU}}

\newcommand{\yr}{\,{\rm yr}}
\newcommand{\myr}{\,{\rm Myr}}

\newcommand{\pc}{\,{\rm pc}}

\newcommand{\msun}{\,M_\odot}

\newcommand{\be}{\begin{equation}}
\newcommand{\ee}{\end{equation}}

\newcommand{\bea}{\begin{eqnarray}}
\newcommand{\eea}{\end{eqnarray}}

\renewcommand{\arcsec}{^{\prime\prime}}

\renewcommand{\farcs}{.\!\!^{\prime\prime}}

\newcommand{\ben}{\begin{enumerate}}
\newcommand{\een}{\end{enumerate}}

\newcommand{\nav}{n_h}

\newcommand{\vsp}{\vspace{2.5mm}}

\begin{document}

\shorttitle{GIANT PLANETS IN BINARIES} 
\shortauthors{PFAHL \& MUTERSPAUGH}


\submitted{Draft 06/25/06}

\title{Impact of Stellar Dynamics on the Frequency of Giant
Planets in Close Binaries}

\author{Eric Pfahl\altaffilmark{1} and
  Matthew Muterspaugh\altaffilmark{2,3}}

\altaffiltext{1}{Kavli Institute for Theoretical Physics, University
of California, Santa Barbara, CA 93106; pfahl@kitp.ucsb.edu}
\altaffiltext{2}{Department of Geological and Planetary Sciences,
California Institute of Technology, Pasadena, CA 91125}
\altaffiltext{3}{Visiting Scientist, Department of Earth, Atmospheric,
and Planetary Sciences, Massachusetts Institute of Technology,
Cambridge, MA, 02141; matthew1@ssl.berkeley.edu}


\begin{abstract}

Hostile tidal forces may inhibit the formation of Jovian planets in
binaries with semimajor axes of $\la$$50\au$, binaries that might be
called ``close'' in this context.  As an alternative to in situ planet
formation, a binary can acquire a giant planet when one of its
original members is replaced in a dynamical interaction with another
star that hosts a planet.  Simple scaling relations for the structure
and evolution of star clusters, coupled with analytic arguments
regarding binary-single and binary-binary scattering, indicate that
dynamical processes can deposit Jovian planets in $<$1\% of close
binaries.  If ongoing and future exoplanet surveys measure a much
larger fraction, it may be that giant planets do somehow form
frequently in such systems.

\end{abstract}




\section{INTRODUCTION}\label{sec:intro}

Surveys using the Doppler technique have identified over 150
extrasolar planets in the last decade.  The available data reveal
important clues to the formation of giant planets around {\em single}
stars \citep[e.g.,][]{Marcy2005}.  Comparatively little is known about
the population of {\em binary} star systems that harbor planets.  Past
planet searches have largely excluded known binaries with angular
separations of $\la$$1\arcsec$, where blending of the two stellar
spectra decreases the sensitivity to small velocity shifts.  Roughly
30 planets have been detected around stars in binaries
\citep{Raghavan2006}.  Most of these binaries are very wide, although
several have separations of $\la$20\,AU, small enough to challenge
standard ideas on Jovian planet formation.  Many more of these compact
systems must be found before we can draw robust conclusions.

In an ongoing targeted search for planets in close, double-lined,
spectroscopic binaries, \citet{Konacki2005} discovered a ``hot
Jupiter'' orbiting the outlying member of the hierarchical triple star
HD 188753.  The inner binary is sufficiently compact that its
influence on the third star is essentially that of a point mass.  What
is intriguing about this system is that a disk around the planetary
host star would be tidally truncated at a radius of only
$\simeq$1\,AU, perhaps leaving insufficient material to produce a
Jovian-mass planet \citep[][]{Jang-Condell2005}.  Broader questions of
how a binary companion impacts planet formation have been explored in
the literature.

If a protoplanetary disk is tidally truncated at $\la$10\,AU, stirring
by the tidal field may prevent the growth of icy grains and
planetesimals, as well as stabilize the disk against fragmentation
\citep[][]{Nelson2000,Thebault2004,Thebault2006}.  In this case,
neither the core-accretion scenario \citep[e.g.,][]{Lissauer1993} nor
gravitational instability \citep[e.g.,][]{Boss2000} are accessible
modes of giant planet formation.  However, the tidal field might also
trigger fragmentation of a marginally stable disk \citep{Boss2006}.
Whether or not giant planet formation is inhibited in close binaries
remains an open problem.

These uncertainties are circumvented if one member of a close binary
is divorced from its original companion and acquires a new partner
star with a planet in tow.  An example of such an event is an exchange
interaction between a binary and a single star in a cluster
environment.  \citet{Pfahl2005} and \citet{PortegiesZwart2005}
proposed a form of this idea as a solution to the puzzle of HD~188753.
Here we present a more general account of dynamical processes that
deposit giant planets in binaries hostile to planet formation.  We
focus exclusively on encounters that mix a binary and a single star or
two binaries.  Higher order multiples are neglected here, but deserve
further attention in light of HD~188753.

We define ``close binary'' in \S~\ref{sec:closebin}.  An overview of
binary scattering dynamics is given in \S~ \ref{sec:scatter}.  Various
aspects of star clusters are summarized in \S~\ref{sec:clusstat}.
Ingredients from \S\S~\ref{sec:scatter} and \ref{sec:clusstat} are
combined in \S~\ref{sec:binplanfrac} to estimate the frequency of
giant planets in close binaries.  In \S~\ref{sec:observations}, our
results are discussed in the context of current exoplanet surveys.


\section{OPERATIONAL DEFINITION OF A CLOSE BINARY}\label{sec:closebin}

Here ``close binary'' refers to the influence of each star's gravity
on the formation and dynamics of planets around its companion.
Consider a binary with semimajor axis $a$, eccentricity $e$, and
stellar masses $M_1$ and $M_2$, and define $q = M_1/M_2$ and $\mu =
M_2/(M_1 + M_2)$.  A disk around star 1 is tidally truncated at radius
\citep{Pichardo2005}
\vsp
\begin{equation}\label{eq:rtrunc}
R_t = [0.733 \,f_{\rm E}(q) \,\mu^{0.07}] \,a(1-e)^{1.2}~,\vsp
\end{equation}
where $f_{\rm E}(q) = 0.49[0.6 + q^{-2/3}\ln (1+q^{1/3})]^{-1}$ is the
Roche-lobe function of \citet{Eggleton1983}.  When $q = 0.1$--10 the
bracketed quantity in eq.~(\ref{eq:rtrunc}) is $\simeq$$0.15$--$0.36$.
We suppose that giant planets form only if $R_t \ga 5$--10\,AU and
$a(1-e)^{1.2} \ga 20$--40\,AU ($q \sim 1$).  In practice, we define a
close binary by $a < 50\au$.

Whether a planet forms in a binary or arrives there dynamically, it
has a maximum orbital radius $a_{p,{\rm max}}$ around its host star
before it is stripped away.  \citet{Holman1999} mapped the range of
stable planetary orbits as a function of $e$ and
$\mu$.  Their polynomial fit is matched to within 15\% by 
\vsp
\begin{equation}\label{eq:holman}
a_{p,{\rm max}} = [0.7\,f_{\rm E}(q)] \,a (1-e)^{1.2}~,\vsp
\end{equation}
a function inspired by eq.~(\ref{eq:rtrunc}).


\section{FEW-BODY DYNAMICS}\label{sec:scatter}

By assumption, all close binaries would be barren of Jovian planets if
not for exchange interactions in their parent clusters.  Before the
exchange, the third star must be single or part of a binary wide
enough to permit giant planet formation.  Exchange is the most robust
dynamical channel for generating close binaries with
planets. Scattering rarely transforms a wide binary into a close
binary, as discussed below.  The main purpose of this section is to
present exchange cross sections for binary-single and binary-binary
scattering.

In binary-single scattering, a binary with semimajor axis $a$ and
component masses $M_1$ and $M_2$ approaches a single star of mass
$M_3$ with a relative speed $u$ at infinity.  Let $M_{12} = M_1 + M_2$
and $M_{123} = M_1 + M_2 + M_3$; other combinations are defined
analogously.  The total system energy vanishes at a critical relative
speed of $u_c = (G\widetilde{M}/a)^{1/2}$, which has a value of
$\simeq$$3\kms$ when $\widetilde{M} = 1\msun$ and $a = 100\au$, where
$\widetilde{M} = M_1M_2M_{123}/M_3M_{12}$ \citep[e.g.,][]{Hut1983}.
For a close binary in an open cluster, we expect $u \sim 1\kms$ (see
\S~\ref{sec:clusstat}) and $u/u_c < 1$, implying that the total energy
is negative and a binary must remain after the interaction.  If $M_3$
passes within $\sim$$a$ from the binary barycenter, the binary may be
strongly perturbed or have an exchange.  The corresponding cross
section is $\sim$$\Sigma_f = 2\pi a\,G M_{123} \, u^{-2}$, determined
by gravitational focusing.  When the masses are similar, a large
fraction of such interactions result in exchange, and so we write the
binary-single exchange cross section as $\Sigma_{\rm bs} = \eta_{\rm
bs}\Sigma_f$, where typically $\eta_{\rm bs}(M_1,M_2,M_3) \la 1$ (see
below).

Large differences between initial and final binary energies are
suppressed by a probability factor $|\Delta E|^{-9/2}$, as shown
analytically by \citet{Heggie1975} and \citet{Heggie1993}.  If $M_3$
replaces $M_1$, and $a'$ is the new semimajor axis, then we expect
$M_1M_2/a' \sim M_2M_3/a$ and $a'/a \sim M_3/M_1$.  For $M_1 \sim
M_2$, we have $\Sigma_{\rm bs} \sim \Sigma_f
(M_3/M_{13})^{7/2}(M_{123}/M_{12})^{1/3}$, based on the analytic
scaling relations of \citet{Heggie1996}.  Although $a'/a \ll 1$ is
possible when $M_3/M_1 \ll 1$, exchange is inhibited by
$(M_3/M_{13})^{7/2}$.  When $M_3/M_1 \gg 1$, the chances of exchange
and $a'/a \gg 1$ are enhanced, but $M_3/M_1 \gg 1$ is unlikely if
$M_3$ is drawn from a stellar mass function such as $p(M_3) \propto
M_3^{-2.3}$.  Based on these arguments, we neglect the shrinkage of
wide binaries and the expansion of close binaries.

A close binary is more likely to encounter a wide binary with
semimajor axis $a_w >50\au$ than a single star \citep[for overviews of
binary-binary scattering, see][]{Mikkola1983a,Mikkola1984a}.  The
cross section for the two binaries to pass closer than $\sim$$a_w$ has
the focusing form, $\sim$$2\pi a_w G M_{1234} u^{-2}$, when $a_w \la
10^3\au$.  When $a_w/a \simeq 1$ the probability is high for exchange
of one star in the close binary with a star in the wide binary.  As
$a_w/a$ increases, there is a decrease in the relative target area of
the close binary and the fraction of encounters that result in
exchange.  In the limit $a_w/a\gg 1$, this fraction should scale as
$a/a_w$, since a star in the wide system must approach within
$\sim$$a$ of the close binary.  Metastable hierarchical triples, which
ultimately dissolve into a binary and single star
\citep[e.g.,][]{Mardling2001}, often result from binary-binary
scattering, which may enhance the exchange fraction somewhat.  We let
$\Sigma_{\rm bb} = \eta_{\rm bb} 2\pi a GM_{1234}u^{-2}$, where
$\eta_{\rm bb}$ depends on the masses and weakly on $a$.  We expect
$\eta_{\rm bb} \la 5$ typically, but this must be checked numerically.

Observations indicate that giant planets orbit $\sim$10\% of single F,
G, and K stars \citep{Marcy2005}. A similar fraction should apply to
stars captured by close binaries in exchange encounters.  Dynamics of
the stars are usually little affected by a planet, but the planet's
orbit may be disrupted.  Let $a_p$ denote the semimajor axis of the
planetary orbit.  \citet{Hut1985} and \citet{Fregeau2004} estimate a
cross section of $\simeq$$2\Sigma_f (a_p/a)^{0.4}$ for two stars to
pass within a distance $a_p$ during a binary-single interaction with
$M_1 = M_2 = M_3$ and $u/u_c \simeq 1$.  Such an approach typically
causes the planet to be ejected from the system, although there is a
significant probability for it to become bound to the other star
\citep{Fregeau2006}.  If $a_p > 1\au$ there is a fair chance that the
planet will be lost in an exchange encounter \nocite{Laughlin1998}
(see Laughlin \& Adams 1998 for a related discussion).  Even if the
planet survives the few-body interaction, its orbit may not have
long-term stability (see eq.~[\ref{eq:holman}]).  We incorporate the
fraction of stars with planets and the ejection probability by
absorbing an ad hoc, constant factor $f_p \la 0.1$ into $\Sigma_{\rm
bs}$ and $\Sigma_{\rm bb}$ (see \S~\ref{sec:binplanfrac}).


\section{STATISTICS OF CLUSTERS AND THEIR STARS}\label{sec:clusstat}

As many as 90\% of all stars form in clusters with
$\sim$$10^2$--$10^3$ members \citep[e.g.,][]{Lada2003}.  Most clusters
disintegrate in $<$100\,Myr, releasing their stars into the Galaxy.
Within 100\,pc of the Sun, a volume containing most exoplanet
discoveries, there are $\sim$$10^5$ binaries contributed by thousands
of clusters.  We aim to determine the percentage close binaries in
this population that harbor giant planets as a result of dynamics.
Since the solar neighborhood samples many stellar birth sites, our
analysis can utilize the gross statistical properties of clusters,
which we now summarize.  

Infant clusters are embedded in gas and dust that dominate the system
mass \citep{Lada2003}.  Embedded clusters (ECs) are easily disrupted
if the diffuse material is expelled rapidly
\citep[e.g.,][]{Hills1980}.  The EC phase lasts for $\la$5\,Myr and
coincides with the critical growth stages of giant planets
\citep[e.g.,][]{Lissauer1993}.  Only $\la$10\% of ECs survive to
become classical open clusters \citep{Lada2003}, but may lose more
than half of their stars following gas expulsion
\citep[e.g.,][]{Boily2003,Adams2006}.  Open clusters (OCs) are also
subject to destructive processes, as reflected in their low median age
of 200\,Myr and the small fraction ($\simeq$2\%) older than 1\,Gyr
\citep[e.g.,][]{Wielen1985}.

For our purposes, a cluster is adequately described by four
parameters: the number of stars $N$, radius $r_h$ enclosing half of
the cluster mass $M_c$ (gas and stars), mean stellar density $\nav =
3N/8\pi r_h^3$ inside $r_h$, and characteristic stellar speed $\sigma
= (GM_c/r_h)^{1/2}$.  ECs have radii scattered about the trend
$r_h({\rm EC}) \simeq 1\,N_2^{1/2}\pc$ \citep[][]{Adams2006} and
masses of $\simeq$$3N\langle M \rangle$ for a 30\% star formation
efficiency, where $N_2 = N/100$, and $\langle M \rangle \simeq
0.5\msun$ is the mean stellar mass.  We see that $\nav({\rm EC})
\simeq 10\,N_2^{-1/2}\pc^{-3}$ and $\sigma({\rm EC}) \simeq
1\,N_2^{1/4}\kms$. OCs have $r_h({\rm OC}) \simeq 1$--5\,pc with a
weak dependence on $N$ and cluster age.  We use a fixed value of
$r_h({\rm OC}) = 1\pc$, so that $\nav({\rm OC}) \simeq
10\,N_2\pc^{-3}$ and $\sigma({\rm OC}) \simeq 0.5\,N_2^{1/2}\kms$.

The natural unit of time for measuring changes in cluster structure is
the half-mass relaxation time:
\vsp
\begin{equation}\label{eq:trelax}
t_{\rm rh} \sim \left(\frac{r_h^3}{G M_c}\right)^{1/2}
\frac{0.1N}{\ln N} 
\simeq
4\,\left(\frac{r_h}{1\pc}\right)^{3/2}N_2^{1/2}\myr~,\vsp
\end{equation}
where we set $\ln N = 5$.  The EC phase is so short ($\la$$1\,t_{\rm
rh}$) that $r_h({\rm EC})$, $n_h({\rm EC})$, and $\sigma({\rm EC})$
change very little.  An OC dissolves as relaxation drives stars across
its tidal boundary; half of the stars escape in a time $T \sim
100t_{\rm rh}$.  Simulations show $N$ dropping almost linearly,
$N(t)/N(0) = 1- t/2T$, where $N(0)$ is the number just after the EC
phase \citep[e.g.,][]{Terlevich1987,PortegiesZwart2001}.  The function
$T = 100\,N_2^{1/2}(0)\myr$ is consistent with simulations and our
kinematical scalings.  This is an upper limit to the true half-life,
since OC decay is hastened by encounters with molecular clouds
\citep[e.g.,][]{Wielen1985}.  Binaries have little impact on the
evolution of typical open clusters \citep[e.g.,][]{Kroupa1995b},
unlike in dense globular clusters, where binaries can strongly modify
the dynamics of core collapse.

An average of some combination of the above $N$-dependent functions
over the cluster ensemble (see \S~\ref{sec:binplanfrac}) requires the
differential $N$ distribution.  For both ECs and young OCs the
distribution is nearly $p(N) \propto N^{-2}$ for $N \sim 10^2$--$10^3$
\citep[e.g.,][]{Elmegreen1997,Lada2003,Adams2006}.  While the most
massive known ECs have a ${\rm few}\times 10^3$ stars, some old OCs
probably had $\ga$$10^4$ stars initially \citep[e.g.,
M67;][]{Hurley2005}.  As a specific choice, we use the range $N =
10^2$--$10^4$ for both ECs and young OCs.

Stellar multiples in clusters must have proportions similar to those
in the Galactic field, where $\simeq$50\%, $\simeq$10\%, and
$\simeq$5\% of stars are binary, triple, and quadruple, respectively
\citep[e.g.,][]{Duq1991}.  Semimajor axes of field binaries span
$\sim$$10^{-2}$--$10^4\au$ and follow a log-normal distribution with a
mean and variance of $\langle \log a({\rm AU})\rangle \simeq 1.5$ and
$\sigma_{\log a} \simeq 1.5$ \citep[e.g.,][]{Duq1991}.  Over any small
range in $a$, $p(a) \propto a^{-1}$ is a good approximation.  The
fraction of binaries that are close ($a \la 50\au$) is $f_{\rm cb}
\simeq 0.5$.  Cluster binary statistics evolve due to dynamical
encounters, but this is evident mainly for systems with $a \ga
10^3\au$ \citep[e.g.,][]{Kroupa1995a}.


\section{FRACTION OF CLOSE BINARIES WITH GIANT PLANETS}\label{sec:binplanfrac}

We now estimate the fraction of close binaries that acquire giant
planets dynamically in clusters.  First, we compute the rate for a
close binary in a cluster to have a favorable interaction.  Then the
cumulative rate for all close binaries is integrated over the cluster
lifetime.  This number is averaged over all clusters, and the result
is divided by the mean number of close binaries per cluster.  Each
step is detailed below for binary-single scattering.  The
binary-binary calculation is completely analogous and only the final
result is quoted.  Note that we neglect interactions between binaries
and stars in the Galactic disk after a cluster dissolves.

Imagine a close binary moving through a cluster of $N$ stars.  Near
the target binary, singles and binaries have densities $n_s$ and
$n_b$, respectively.  We let $n_s = f_s n$, $n_b = f_b n$, and $f_s +
f_b = 1$, and assume that $f_s$ and $f_b$ are independent of $N$, $t$,
and position within the cluster.  We assume that all objects have a
Maxwellian speed distribution with one-dimensional velocity dispersion
$\sigma$.  Relative speeds $u$ then also follow a Maxwellian
distribution, but with dispersion $\sqrt{2}\sigma$.  The rate for the
the target binary to acquire a single star and its planet is
\vsp
\begin{eqnarray}\label{eq:bsrate}
n_s\langle\Sigma_{\rm bs} u\rangle & = & 
2\sqrt{\pi} n_s f_p a \,G 
\langle \eta_{\rm bs} \,M_{123} \rangle \,\sigma^{-1} \nonumber \\ 
& \simeq & 7.5 \times 10^{-12} 
\,n_1 \,a_2 \, \sigma_0^{-1} 
\left(
\frac{f_p}{0.1}
\frac{\langle \eta_{\rm bs}\,M_{123} \rangle}{M_\odot}
\right) \yr^{-1}
~,\vsp
\end{eqnarray}
where $n_1 = n_s/10\pc^{-3}$, $a_2 = a/100\au$, $\sigma_0 =
\sigma/1\kms$, and the angled brackets denote averages over $u$ and
$M_3$.  If $n_s$ and $\sigma$ take their characteristic values for an
open cluster (see \S~\ref{sec:clusstat}), we find that over the
half-life $T$ the planet-capture probability is $\sim$$10^{-3}N_2
a_2$; such encounters are rare.  Since $n_s\langle\Sigma_{\rm bs}
u\rangle \propto a$, the $a$ distribution for close binaries that do
acquire planets may be nearly flat if the primordial distribution is
$\propto$$a^{-1}$.

Integration of $n_s\langle \Sigma_{\rm bs} u\rangle$ over all close
binaries and the cluster lifetime gives the total number of planet
captures from binary-single exchange encounters:
\vsp
\begin{eqnarray}\label{eq:nbsint}
N_{\rm bs} & = & 
\int dt \int dV \, n_b\,n_s 
\langle\langle \Sigma_{\rm bs} u \rangle\rangle \nonumber \\
& = &
2\sqrt{\pi}
f_s\,f_b\,f_p\,
\,G \langle\langle a \eta_{\rm bs} M_{123} \rangle\rangle
\int dt \sigma^{-1}
\int dV n^2~,\vsp
\end{eqnarray}
where $dV$ is a volume element, $n_b$ is the local number density of
binaries (close and wide), and double brackets denote averages over
$u$, the $M_i$, and $a$ for the close binaries.  Among close binaries,
the mean $a$ is $\sim$10\,AU.  The volume integral picks out the
formal mean density: $\int dV n^2 = \int dN\,n \equiv N \langle
n\rangle$.  Density profiles appropriate for open clusters have
$\langle n \rangle \simeq n_h$; we equate these two densities.

The approximate scaling relations in \S~\ref{sec:clusstat} allow us to
evaluate $N_{\rm bs}$ for ECs and OCs.  We assume that the EC
phase lasts $10^7\yr$ and has fixed $N$, which gives
\vsp
\begin{equation}\label{eq:nbsec}
N_{\rm bs}({\rm EC}) \simeq 0.0002
\,N_2^{1/4}
\left(
\frac{f_s}{0.5}
\frac{f_b}{0.5}
\frac{f_p}{0.1}
\frac{\langle\langle a\,\eta_{\rm bs} \,M_{123} \rangle\rangle}{10\au\msun}
\right)
~.\vsp
\end{equation}
If the number of open-cluster stars drops linearly in time (see
\S~\ref{sec:clusstat}), integration over the full lifetime $2T$ gives
\vsp
\begin{equation}\label{eq:nbsoc}
N_{\rm bs}({\rm OC}) \simeq 0.003\,N_2^{2}
\left(
\frac{f_s}{0.5}
\frac{f_b}{0.5}
\frac{f_p}{0.1}
\frac{\langle\langle a\,\eta_{\rm bs}\,M_{123} \rangle\rangle}{10\au\msun}
\right)
~,\vsp
\end{equation}
where $N_2 = N(0)/100$.  For the fraction $f_{\rm oc} \la 0.1$ of
newly minted clusters that are destined to be open, the early embedded
phase yields only a small correction to $N_{\rm bs}$.

Each cluster disperses $f_b f_{\rm cb} N$ close binaries into the
Galaxy.  Using $p(N)$ in \S~\ref{sec:clusstat}, we sum $N_{\rm bs}$
over an ensemble of clusters and divide by the total number of close
binaries to obtain the fraction of all close binaries that acquire a
planet via binary-single exchange:
\vsp
\begin{eqnarray}
F_{\rm bs} & \simeq & 
\frac{f_{\rm oc}\langle N_{\rm bs}({\rm OC})\rangle + 
(1-f_{\rm oc})\langle N_{\rm bs}({\rm EC}) \rangle}
{f_b f_{\rm cb} \langle N\rangle} \nonumber \\
& \simeq & 
0.0003
\left(
\frac{f_{\rm oc}}{0.1}
\frac{0.5}{f_{\rm cb}}
\frac{f_s}{0.5}
\frac{f_p}{0.1}
\frac{\langle\langle a\,\eta_{\rm bs} \,M_{123} \rangle\rangle}{10\au\msun}
\right)
~,\vsp
\end{eqnarray}
where the overall contribution from ECs is negligible.  For
binary-binary scattering, we replace $n_s$ and $\Sigma_{\rm bs}$ with
$n_b$ and $\Sigma_{\rm bb}$ in eq.~(\ref{eq:nbsint}), and estimate
\vsp
\begin{equation}
F_{\rm bb} \simeq 
0.0003
\left(
\frac{f_{\rm oc}}{0.1}
\frac{0.5}{f_{\rm cb}}
\frac{f_b}{0.5}
\frac{f_p}{0.1}
\frac{\langle\langle a\,\eta_{\rm bb} \,M_{1234} \rangle\rangle}{10\au\msun}
\right)
~,\vsp
\end{equation}
where we expect $\langle\langle a\,\eta_{\rm bb} \,M_{1234}
\rangle\rangle \la 100\au\msun$ (see \S~\ref{sec:scatter}).

The fraction of all close binaries that capture giant planets is
$F_{\rm ex} = F_{\rm bs} + F_{\rm bb} \la 10^{-3}$ if the above
parameters take their plausible fiducial values.  Reasonable
variations in our adopted scaling relations or more accurate cross
sections might yield $F_{\rm ex}\sim 10^{-2}$.  Small values of
$F_{\rm ex}$ result from the relative rarity of suitable exchange
encounters in open clusters (see the text below
eq.~[\ref{eq:bsrate}]).  For future theoretical work, we recommend a
systematic study of few-body interactions including planets in order
to obtain better cross sections.  Complementary $N$-body simulations
of clusters with binaries and planets would stringently test of our
assertions.


\section{COMPARISON TO OBSERVATIONS}\label{sec:observations}

Raghavan et al. (2006; see also Eggenberger et al. 2004)
\nocite{Raghavan2006,Eggenberger2004} find that $\simeq$30 of
$\simeq$130 exoplanet host stars have binary companions, most with
separations of $10^2$--$10^4\au$.  Only 5 systems (see Table~1) are
candidate close binaries; 2 are technically triples.  The objects in
Table~1 were observed in different surveys, each with different
criteria to select targets.  This makes it difficult to empirically
estimate the fraction, $F$, of close binaries with giant planets.
Given that $\simeq$3000 stars have been searched for planets, the vast
majority of which are not close binaries, the expectation is that $F
\gg 0.1\%$.

\input{tab1.tex}
Known spectroscopic binaries with angular separations of
$\la$1--$2\arcsec$ have been largely overlooked in Doppler surveys.
HD~41004, HD~196885, GJ~86, and $\gamma$ Cephei have relatively large
angular separations of $\simeq$$0\farcs 5$, $\simeq$$0\farcs 7$,
$\simeq$$1\arcsec$, and $2\arcsec$, respectively.  Perhaps more
importantly, the secondary stars in these systems are sufficiently
faint that the primary's spectrum is not greatly contaminated.  The
serious selection effects against discovering giant planets in close
binaries strengthen the notion that $F \gg 0.1\%$.

We note that $F \sim 1\%$ is consistent with the discovery of a planet
in HD~188753 by \citet{Konacki2005}, who has so far conducted a
cursory analysis of $\simeq$100 binaries.  A similar fraction follows
from the limited \citet{Campbell1988} survey of 16 stars that
ultimately led to the detection of a planet in $\gamma$ Cephei
\citep{Hatzes2003}.  Our preferred value for the contribution to $F$
from dynamics is $\sim$0.1\%.  If future surveys verify that $F\sim
1\%$, this may signal that giant planets do form in close binaries,
despite the seemingly unfavorable conditions.  

Several exoplanet searches that specifically target close binaries are
now underway.  Konacki's spectroscopic survey includes $\simeq$100
known binaries with projected separations of 10--60\,AU and distances
of 30--$300\pc$.  The Palomar High-precision Astrometric Search for
Exoplanet Systems \citep[PHASES; ][]{LaneMute2004a, Muterspaugh2005}
at the Palomar Testbed Inteferometer \citep{Colavita1999} aims to
monitor $\simeq$50 visual binaries, 16 of which also belong to
Konacki's radial-velocity sample. \cite{Udry2004} report on the first
steps in a campaign to spectroscopically search for planets in
$\simeq$100 single-line spectroscopic binaries with 2-50 year periods
($a \simeq 2$--$15\au$).


\acknowledgements

We thank Phil Arras and the referee, John Chambers, for valuable
comments, and Maciej Konacki for providing details on his survey.  EP
was supported by NSF grant PHY99-07949.  MWM was supported by the
Michelson Graduate Fellowship and NASA grant NNG05GJ58G issued through
the Terrestrial Planet Finder Foundation Science Program.  MWM
appreciates the hospitality of the KITP during the week in which this
work began.




\end{document}

%% file: tab1.tex
\begin{deluxetable}{p{2.2cm}lcccc}
\tablecolumns{6}
\tablewidth{3.5in}
\tablecaption{Close Binaries with Planets. \label{tab:tabclose}}
\tabletypesize{\scriptsize}
\tablehead{
\colhead{Object} &
\colhead{$a$(AU)} &
\colhead{$e$\tablenotemark{a}} &
\colhead{$M_1/M_2$\tablenotemark{b}} &
\colhead{$R_t$(AU)} &
\colhead{Refs}
}
\startdata
HD 188753\tablenotemark{c} \dotfill & 12.3 & 0.50 & 1.06/1.63 & 1.3 & 1 \\
$\gamma$ Cephei \dotfill & 18.5 & 0.36 & 1.59/0.34 & 3.6 & 2,3 \\
GJ 86 \tablenotemark{d} \dotfill & $\sim$20 & \nodata & 0.7/1.0 & $\sim$5 & 4,5,6 \\
HD 41004\tablenotemark{e} \dotfill & $\sim$20 & \nodata & 0.7/0.4 & $\sim$6 & 7 \\
HD 196885 \dotfill & $\sim$25 & \nodata & 1.3/0.6 & $\sim$7 & \phn\phn\phn8 
\enddata
\tablenotetext{a}{When no eccentricity is given, only the projected binary separation is known.}
\tablenotetext{b}{Planetary host mass divided by companion mass.}
\tablenotetext{c}{The secondary is a binary with semimajor axis 0.67\,AU.}
\tablenotetext{d}{The secondary is a white dwarf of mass $\simeq$$0.5\msun$.  
To estimate $R_t$, we assumed an orginal companion mass of $1\msun$.}
\tablenotetext{e}{The secondary is orbited by a brown dwarf with a 1.3\,day period.}
\tablerefs{
(1) Konacki 2005; \nocite{Konacki2005};
(2) Campbell et al. 1988; \nocite{Campbell1988}
(3) Hatzes et al. 2003; \nocite{Hatzes2003};
(4) Queloz et al. 2000; \nocite{Queloz2000} 
(5) Mugrauer \& Neuh{\"a}user 2005; \nocite{Mugrauer2005} 
(6) Lagrange et al. 2006; \nocite{Lagrange2006}
(7) Zucker et al. 2004; \nocite{Zucker2004} 
(8) Chauvin et al. 2006 \nocite{Chauvin2006}
}
\end{deluxetable}